\documentclass[10pt,a4paper]{article}
\usepackage{amsmath,amsfonts,amssymb,bbm,citesort}

\newcommand{\eqn}[1]{\begin{equation} #1 \end{equation}} % equation environment
\newcommand{\aln}[1]{\begin{align} #1 \end{align}}       % align environment
\newcommand{\mul}[1]{\begin{multline} #1 \end{multline}} % multline environment
\newcommand{\mc}{\mathcal}                               % script letters
\newcommand{\msf}{\mathsf}                               % straight letters
\newcommand{\mbf}{\mathbf}                               % bold letters
\newcommand{\eq}[1]{(\ref{#1})}              % equation reference
\newcommand{\pd}{\partial}                   % partial derivative sign
\newcommand{\wtilde}{\widetilde}             % wide tilde
\newcommand{\wbar}{\overline}                % wide bar
\renewcommand{\l}{\left}
\renewcommand{\r}{\right}

\begin{document}

\title{Semiclassical theory in Andreev billiards:\\ 
       beyond the diagonal approximation}

\author{Oleg Zaitsev\\[.3cm]
\emph{Fachbereich Physik, Universit\"at Duisburg-Essen,}\\ 
\emph{45117 Essen, Germany} \\[.3cm]
E-mail: \texttt{oleg.zaitsev@uni-duisburg-essen.de}}
\date{}

\maketitle

\begin{abstract}

Recently semiclassical approximations have been successfully applied to study
the effect of a superconducting lead on the density of states and conductance
in ballistic billiards. However, the summation over classical trajectories
involved in such theories was carried out using the intuitive picture of
Andreev reflection rather than the semiclassical reasoning. We propose a method
to calculate the semiclassical sums which allows us to go beyond the diagonal
approximation in these problems. In particular, we address the question of
whether the off-diagonal corrections could explain the gap in the density of
states of a chaotic Andreev billiard. \vspace{.5cm}

PACS: 05.45.Mt, 74.45.+c

\end{abstract}

\section{Introduction}

Andreev billiards are two-dimensional electron cavities with a superconducting
part of the boundary (see \cite{been05} for a review). A negatively charged
electron incident on the superconductor is reflected as a positively charged
hole with an opposite momentum. This process of Andreev
reflection~\cite{andr64} generates a new kind of dynamics in comparison to
conventional (normal) billiards. The excitation spectrum of an Andreev billiard
depends on the shape of its boundary~\cite{mels96,ihra01}. Using the
random-matrix theory, it was shown~\cite{mels96} that the density of
states (DoS)~$d(E)$ in the chaotic billiard has a minigap around the Fermi
energy~$E_F$, while in the integrable billiard it is proportional to~$E$, the
energy counted from~$E_F$. The width of the minigap is of the order of the
Thouless energy, which is much smaller than the superconducting gap~$\Delta$.
The semiclassical theory of~\cite{ihra01} confirms the above results in the
integrable case, but finds an exponential suppression of $d(E)$, instead of a
gap, in the chaotic cavity. The authors surmised that the disagreement can be
attributed to the use of diagonal approximation. The method proposed below
allows one to determine the off-diagonal corrections in a billiard with a
superconducting lead. (A somewhat similar concept was used in~\cite{adag02} in
the case where the whole boundary is superconducting.) We find that these
corrections may, indeed, reflect the existence of the gap, although the full
understanding of the problem is still missing. 

Another area of the application of the semiclassical techniques is transport
problems. In general, the conductance of a billiard with a superconducting lead
can be expressed in terms of transmission and reflection amplitudes with or
without electron-to-hole conversion~\cite{taka92,lamb93}. In particular, the
contribution of the Andreev reflection to the conductance is proportional to
$R_A \equiv \text{Tr}\, (r_{he} r_{he}^\dag)$, where $r_{he}$ is a part of the
scattering matrix describing an electron injected into the billiard returning
back as a hole. In~\cite{lass03}, this quantity was calculated semiclassically
under the assumption of the exact Andreev reflection ($E \to 0$) for a
three-probe geometry. Both the diagonal and off-diagonal (weak localisation)
contributions were given. It was argued there that $R_A$ is equal simply to the
normal transmission coefficient for an electron to reach the superconducting
lead (then, classically, it would return back as a hole with a 100\%
probability). Hence, the off-diagonal corrections to $R_A$ could be computed by
standard methods~\cite{rich02} developed for a normal billiard. In this letter,
we start from the \emph{quantum} relation~\cite{been92} between $R_A$ and the
normal transmission amplitudes (in the case of a two-probe geometry) and then
apply the semiclassical limit. Thus, no additional classical assumption is
required. It is found that, while the diagonal part of $R_A$ is, in fact, equal
to the classical normal transmission, the weak-localisation correction is
nontrivial.

\section{Density of states}

\subsection{Problems with the earlier approach}
\label{prob}

Here, we outline the semiclassical theory of~\cite{ihra01} and point out the
technical difficulties that arise when sums over classical trajectories are
calculated. We consider a chaotic billiard with a superconducting lead. The
lead width $w$ is assumed to be small compared to the perimeter of the
billiard, so as not to spoil its chaoticity. The energy is taken to be high
enough to provide for a large number of open channels $N = k_F w/ \pi \gg 1$ in
the lead ($k_F$ is the Fermi wave number). The lead is modelled as an ideal
wire connecting the billiard and the superconductor. The average quasiparticle
DoS is given by~\cite{ihra01}
\eqn{
  d(E) = d_0 - \frac 1 \pi \text{Im} \sum_{l=1}^\infty \frac {(-1)^l} l \frac
  \pd {\pd E} \l< \text{Tr}\,[S(E) S^* (-E)]^l \r>,
\label{dE}
}
where $d_0 = M \msf A/ 2 \pi \hbar^2$ is the average DoS for a particle
of mass~$M$ in the isolated billiard of area~$\msf A$, $S(E)$~is the electron
scattering matrix for the billiard with the normal lead, the energy~$E$ is
measured from~$E_F$ and the average is taken over a classically small interval
of~$k_F$. In the semiclassical approximation, the scattering
matrix~\cite{mill74,doro92} 
\eqn{
  S_{nm} (E) = \sum_{\gamma (n,m)} \mc A_\gamma \exp \l( \frac i \hbar \mc
  S_\gamma \r) 
\label{Ssem}
}
is a sum over classical trajectories $\gamma (n,m)$ starting and ending at the
lead and connecting the channels~$m$ and~$n$. The trajectories make an angle
$\theta_m = \pm \sin^{-1} (m \pi /k_F w)$ with the lead direction as they enter
the billiard, and $\theta_n $---as they leave it.
Here we assumed $|E| \ll E_F$. The actions then can be expanded about $E_F$ as
$\mc S_\gamma (E) \simeq \mc S_\gamma (0) + E \mc T_\gamma (0)$, where $\mc
T_\gamma$ is the time and
\eqn{
  \mc S_\gamma (p_n, p_m; E=0) = \int_\gamma \mbf p \cdot d \mbf r - p_n y_n +
  p_m y_m
}
is in the momentum representation on the lead. Here, $y_{m(n)}$ are initial
(final) coordinates at the lead cross section and $p_{m(n)} = \hbar k_F \sin
\theta_{m(n)}$ are the conjugate momenta. The prefactors $\mc A_\gamma = |\mc
A_\gamma| \exp(-i \mu_\gamma \pi /2)$ depend on the the Maslov
indices~$\mu_\gamma$ and
\eqn{
  |\mc A_\gamma| = \l( \frac \pi {2 k_F w^2} \l| \frac {\pd y_n} {\pd
  (\sin \theta_m)} \r| \r)^{1/2} = \l( \frac {\pi \hbar} {2 w^2} \l| \frac
  {\pd^2 S_\gamma } {\pd p_n \pd p_m} \r| \r)^{1/2}.
\label{A2}
}

Using the semiclassical expression~\eq{Ssem}, the authors of~\cite{ihra01}
arrive, in the diagonal approximation, at the following result:
\aln{
  \l< \text{Tr}\, [S(E) S^*(-E)]^l \r> &= \sum_{n,m = 1}^N \, \sum_{\gamma
  (n,m)} \l|\mc A_\gamma \r|^2 \exp \l( \frac i \hbar 2 l E \mc T_\gamma  \r)
  \notag \\
  &\simeq \frac {k_F} {2 \pi} \int _{-1}^1 d (\sin \theta) \int_0^w dy \exp \l[
  \frac i \hbar 2 l E \mc T (y, \theta) \r],
\label{TrS}
} 
where the sums become the integrals over the initial (or final) conditions $(y,
\theta)$ in the limit $N \gg 1$. While the derivation of this formula for $l=1$
is given in~\cite{ihra01}, its generalization to $l>1$ is not obvious. For
example, when $l=2$, we can write
\aln{
  &\text{Tr}\, [S(E) S^*(-E)]^2 \notag \\
  &= \sum_{\substack{n_1, n'_1,\\ n_2, n'_2 = 1}}^N\;\;
  \sum_{\substack{\gamma_1 (n_1, n'_1), \gamma'_1 (n'_1, n_2),\\ \gamma_2 (n_2,
  n'_2), \gamma'_2 (n'_2, n_1)}} \mc A_{\gamma_1} \mc A_{\gamma'_1}^* \mc
  A_{\gamma_2} \mc A_{\gamma'_2}^* \exp \l[ \frac i \hbar \l(\mc S_{\gamma_1} -
  \mc S_{\gamma'_1} + \mc S_{\gamma_2} - \mc S_{\gamma'_2} \r) \r] \notag \\
  &\times  \exp \l[ \frac i \hbar E \l(\mc T_{\gamma_1} + \mc T_{\gamma'_1} +
  \mc T_{\gamma_2} + \mc T_{\gamma'_2} \r) \r].
\label{suml2}
}
After this expression is averaged over a small window of~$k_F$, only the terms
where the actions cancel in the exponent will remain. The diagonal
approximation amounts to keeping the terms with all four actions equal. In a
chaotic system without symmetries this is only possible when $\gamma_1 =
(\gamma'_1)^{-1} = \gamma_2 = (\gamma'_2)^{-1}$, where $\gamma^{-1}$ is the
time reversed path~$\gamma$. This condition corresponds to the exact Andreev
reflection at $E = 0$. It yields the average 
\eqn{
  \l< \text{Tr}\, [S(E) S^*(-E)]^2 \r> = \sum_{n, n'} \; \sum_{\gamma (n, n')}
  |\mc A_\gamma|^4 \exp \l( \frac i \hbar 4 E \mc T_\gamma \r),
}
which is a wrong result. Thus, this calculation shows that the diagonality
requirement alone is not sufficient in the case of Andreev billiards. As we
shall see next, the stationary-phase (SP) approximation must be employed when
calculating the semiclassical sums.

\subsection{Traces of scattering matrices}

In this section, we derive equation~\eq{TrS} and find the quantum
corrections to it. In order to focus on the idea of the method, the case $l=2$
is considered here and the general situation is treated in appendix~\ref{genl}.

\subsubsection{Classical contribution}

We begin by rewriting the channel sum~\eq{suml2} as an integral over momenta
\mul{
  \text{Tr}\, [S(E) S^*(-E)]^2 = \l(\frac w {\pi \hbar} \r)^4 \int_{-\hbar
  k_F}^{\hbar k_F} dp_1 dp'_1 dp_2 dp'_2 \sum_{\substack{\gamma_1, \gamma'_1,\\
  \gamma_2,\gamma'_2}} \mc A_{\gamma_1} \mc A_{\gamma'_1}^* \mc A_{\gamma_2}
  \mc A_{\gamma'_2}^*  \\  
  \times \exp \l[ \frac i \hbar \l(\mc S_{\gamma_1} -
  \mc S_{\gamma'_1} + \mc S_{\gamma_2} - \mc S_{\gamma'_2} \r) \r] 
  \exp \l[ \frac i \hbar E \l(\mc T_{\gamma_1} + \mc T_{\gamma'_1} +
  \mc T_{\gamma_2} + \mc T_{\gamma'_2} \r) \r].
\label{intl2}
}
Although each channel allows for two signs of the momentum, it is expected that
the leading contribution to the integral comes from the Andreev-reflected
paths. Therefore, given $p_1$, $p'_1$, $p_2$ and $p'_2$, the initial and final
momenta of the four paths are set as follows: $\gamma_1 (p_1 \leftarrow
-p'_1)$, $\gamma'_1 (p'_1 \leftarrow -p_2)$, $\gamma_2 (p_2 \leftarrow -p'_2)$
and $\gamma'_2 (p'_2 \leftarrow -p_1)$. 

It is convenient to compute the integral~\eq{intl2} in the rotated coordinates 
\eqn{
  \l\{
  \begin{array}{l}
    a = \frac {p_1 - p_2} {\sqrt 2} \\
    b = \frac {p_1 + p_2} {\sqrt 2}
  \end{array}
  \r. \quad \text{and} \quad \l\{
  \begin{array}{l}
    a' = \frac {p'_1 - p'_2} {\sqrt 2} \\
    b' = \frac {p'_1 + p'_2} {\sqrt 2}
  \end{array}
  \r. .
\label{trans2}
}
It will be possible to integrate over $da\, da'$ in the SP approximation. 
Consider the phase function
\eqn{
  \Phi (a, b, a', b') \equiv \mc S_{\gamma_1} (p_1, -p'_1) - \mc S_{\gamma'_1}
  (p'_1, -p_2) + \mc S_{\gamma_2} (p_2, -p'_2) - \mc S_{\gamma'_2} (p'_2,
  -p_1),
}
where the momenta must be expressed in terms of the new coordinates. It is easy
to show that the SP condition $\pd \Phi / \pd a = \pd \Phi / \pd a' = 0$ can be
rewritten as a relation between the initial and final positions~$y^{i,f}$ of
the trajectories on the lead:
\eqn{
  -y_{\gamma_1}^f - y_{\gamma'_1}^i + y_{\gamma_2}^f + y_{\gamma'_2}^i =
  -y_{\gamma_1}^i + y_{\gamma'_1}^f + y_{\gamma_2}^i - y_{\gamma'_2}^f = 0.
}
These equations are satisfied if 
\eqn{
  \gamma_1 = \gamma_2 \quad \text{and} \quad \gamma'_1 = \gamma'_2, 
\label{sttr}
}
and the stationary point is given by $a = a' = 0$. Now it is necessary to
employ the diagonal-approximation requirement $\Phi = 0$. It limits possible
trajectories to $\gamma_1 = (\gamma'_1)^{-1} = \gamma_2 = (\gamma'_2)^{-1}$, as
expected. Since the condition $\Phi = 0$ is fulfilled for arbitrary $b$
and~$b'$, the integral over $db\, db'$ must be performed exactly. In the
original coordinates, it is the integral over the manifold 
\eqn{
  \l\{
  \begin{array}{l}
    p_1 = p_2 \\
    p'_1 = p'_2
  \end{array}
  \r.  \quad (|p_i|, |p'_i| \leq k_F).
\label{SPman}
}

To complete the program, we compute the second derivatives of the phase at the
SP point:
\eqn{
  \left. \frac {\pd^2 \Phi} {\pd a^2} \right|_{a = a' = 0} = \left. \frac
  {\pd^2 \Phi} {\pd a'^2} \right|_{a = a' = 0} = 0,  \quad \quad
  \left. \frac {\pd^2 \Phi} {\pd a \pd a'} \right|_{a = a' = 0} = 2 \frac
  {\pd^2 \mc S_{\gamma_1} (p_1, -p'_1)} {\pd p_1\, \pd (-p'_1)}.
}
The SP integration leads to the cancellation of $|\mc A_{\gamma_1}|^2$
in~\eq{intl2}, according to~\eq{A2}. Finally, transforming the $db\, db'$
integral to the $dp_1\, dp'_1$ integral, we arrive at~\eq{TrS}.

It is instructive to consider an analogy with the semiclassical treatment of
normal reflection. Suppose, the scattering matrix $S_1$ ($S_2$) describes the
propagation of the particle before (after) the reflection and $S = S_2 S_1$ is
the total scattering matrix. Semiclassically, $S_1$ and $S_2$ are given as sums
over classical trajectories. If $S$ is calculated in the SP approximation, it
will be a sum over combined classical trajectories which include the
reflection. In the case of Andreev billiard, the consequence of the SP
approximation is the equality of trajectories~\eq{sttr}, while the Andreev
reflection is required by the diagonal approximation.

\subsubsection{Quantum corrections}

There are two types of quantum corrections to~\eq{TrS}: diagonal and
off-diagonal. (The situation here is analogous to the calculation of the
quantum corrections to reflection in a normal billiard~\cite{rich02}.) The
off-diagonal corrections result from the
Sieber-Richter~(SR)~\cite{sieb01,rich02} pairs of trajectories. Such
trajectories are exponentially close in the phase space (up to a time
reversal), apart from a small region where one of them has a self-crossing and
the other has an anticrossing. Their actions differ by a small amount~$\Delta
\mc S (\varepsilon)$ depending on the crossing angle~$\varepsilon$. Without the
diagonality requirement $\Phi = 0$, a more general condition~\eq{sttr} should
be used. If $\gamma_1 = \gamma_2$ has a crossing (anticrossing), it can be
paired with $\gamma'_1 = \gamma'_2$ which begins and ends exponentially close
to $(\gamma_1)^{-1}$, but has an anticrossing (crossing). Their phase
difference $\Phi = \pm l \Delta \mc S (\varepsilon_{\gamma_1})$ (see
appendix~\ref{genl} for $l>2$) is small compared to their actions. Hence, the
momentum integration, as above, yields the SR correction to~\eq{TrS} 
\mul{
  \l< \text{Tr}\, [S(E) S^*(-E)]^l \r>_{\text{SR}} =\\ 
  2 \sum_{n,m = 1}^N \, \sum_{\{\gamma (n,m)\}} \l|\mc A_\gamma \r|^2 \exp \l(
  \frac i \hbar 2 l E \mc T_\gamma  \r) \cos \l[ \frac {l \Delta \mc S
  \l(\varepsilon_{\{\gamma\}}\r)} \hbar \r],
\label{TrSSR1}
}
where the index $\{\gamma\}$ runs over all self-crossings of the path~$\gamma$.
Performing the summation over the crossings by the standard
procedure~\cite{sieb01,rich02,zait05b,heus06}, we arrive at the
final result
\eqn{
  \l< \text{Tr}\, [S(E) S^*(-E)]^l \r>_{\text{SR}} = - \frac 1 {l N \mc
  T_{\text{esc}}} \sum_{n,m = 1}^N \, \sum_{\gamma (n,m)} \l|\mc A_\gamma \r|^2
  \mc T_\gamma \exp \l( \frac i \hbar 2 l E \mc T_\gamma  \r)
\label{TrSSR}
}
depending on the average escape time~$\mc T_{\text{esc}} = \pi \msf A/w v_F$,
where $v_F$ is the Fermi velocity (appendix~\ref{SR}). 

The diagonal quantum correction is
\eqn{
  \l< \text{Tr}\, [S(E) S^*(-E)]^l \r>_{\text{diag}} = \frac 1 l \sum_{n=
  1}^N \, \sum_{\gamma (n,n)} \l|\mc A_\gamma \r|^2 \exp \l( \frac i \hbar 2 l
  E \mc T_\gamma  \r).
\label{TrSd}
} 
For $l=1$, it is readily derived using equation~\eq{Ssem}. Namely, in $\l<
\text{Tr}\, S S^*\r> = \sum_{nm} \sum_{\gamma, \gamma'} \l< \cdots \r>$ the
terms with $n=m$ and $\gamma = \gamma'$ are considered. Since the actions of
the \emph{identical} paths cancel in the phase, these terms survive the
averaging over~$k_F$ and yield the above result. Note that the terms with the
\emph{time-reversed} paths $\gamma = \gamma'^{-1}$ enter the classical
part~\eq{TrS}. In the case \mbox{$l>1$}, the preceding integration procedure
needs to be modified. This is done in appendix~\ref{diagcor}. It is worth
mentioning that both the SR and diagonal corrections are of the next order in
$N^{-1}$ to the classical contribution.

For the following calculation, it will be helpful to have the sums over the
trajectories in \eq{TrSSR} and~\eq{TrSd} transformed into integrals over time.
This can be achieved by using the sum rule for chaotic billiards~\cite{rich02}:
\eqn{
  \sum_{\gamma (n,m)} \l|\mc A_\gamma \r|^2 \delta (\mc T - \mc T_\gamma)
  \simeq N^{-1} P (\mc T),
\label{sumr}
}
where $P (\mc T) = (\mc T_{\text{esc}})^{-1} \exp (- \mc T/ \mc
T_{\text{esc}})$ is the survival probability. Adding the SR and diagonal
contributions together, we obtain the total quantum correction
\eqn{
  \l< \text{Tr}\, [S(E) S^*(-E)]^l \r>_{\text{quant}} = \frac 1 l \int_0^\infty
  d \mc T P (\mc T) \l(1 - \frac {\mc T} {\mc T_{\text{esc}}} \r) \exp \l(
  \frac i \hbar 2 l E \mc T  \r).
\label{TrSq}
}

\subsection{Quantum correction to the density of states}

The quantum correction to the DoS, originating from the SR and diagonal
contributions, 
\eqn{
  \delta d(E) = \frac 2 {\pi \hbar} \int_0^\infty d \mc T P (\mc T) \mc T \l(1
  - \frac {\mc T} {\mc T_{\text{esc}}} \r) \ln \l(2 \l|\cos \frac {E \mc T}
  \hbar  \r| \r),
\label{dd}
}
is found by substituting the corrections to the traces~\eq{TrSq} in
equation~\eq{dE}. In deriving~\eq{dd}, the sum over $l$ was computed as
follows:
\aln{
  \text{Re} \sum_{l=1}^\infty \frac {(-1)^l} l \exp \l( \frac i \hbar 2 l E \mc
  T  \r)  &= - \text{Re}\, \ln \l[1 + \exp \l( \frac i \hbar 2 E \mc T  \r) \r]
  \notag\\
  &= - \ln \l(2 \l|\cos \frac {E \mc T} \hbar  \r| \r).
}

\begin{sloppypar}
Important conclusions can be drawn from~\eq{dd} already in the limit \mbox{$E
/E_{\text{Th}} \ll 1$}, where $E_{\text{Th}} = \hbar / 2 \mc T_{\text{esc}}$ is
the Thouless energy. In this case, one finds
\eqn{
  \delta d(E) \approx - \frac 1 {\pi E_{\text{Th}}} \l[ \ln 2 - \l( \frac 3 2
  \frac E {E_{\text{Th}}} \r)^2 \r].
\label{ddlim}
}
We recall that the classical part of the DoS~\cite{scho99,ihra01} 
\eqn{
  d_{\text{cl}}(E) = d_0 x^2 \frac {\cosh x} {\sinh^2 x}, \quad x \equiv \frac
  {\pi E_{\text{Th}}} E,
}
becomes exponentially small in this limit. Hence, equation~\eq{ddlim} implies
that the total DoS is \emph{negative} at small energies. This unphysical result
has two possible explanations: either there are other sources of quantum
corrections not taken into account in the present work or the semiclassical
theory becomes inapplicable near~$E_F$, thereby reflecting the existence of the
gap. Assuming the latter, we can roughly estimate the gap width~$E_g$ by
setting $d_{\text{cl}}(E_g) + \delta d(E_g) = 0$. This energy is weakly
dependent on the channel number~$N$ and reads $E_g/E_{\text{Th}} \approx 0.45,
\, 0.34$ for $N = 10,\, 100$, respectively. These numbers are comparable with
the results of the random-matrix theory~\cite{mels96} $E_g/E_{\text{Th}}
\approx 0.6$, as well as the full quantum calculations~\cite{ihra01}.
\end{sloppypar}

\section{Conductance}

In this section, we derive semiclassical formulae for the conductance~$G$ of a
chaotic billiard having one normal (N) and one superconducting (S) lead. It was
shown~\cite{taka92,lamb93} that $G = (4 e^2/ h)\, R_A$, where $R_A$ is the
Andreev-reflection coefficient defined in the introduction. It can be
expressed~\cite{been92} in terms of the electron transmission matrices
$t_{\text{SN}}$ and $t_{\text{NS}}$ between the leads~as
\eqn{
  R_A = \text{Tr}\, \l[t_{\text{SN}} t_{\text{NS}}^* \l(2 - t_{\text{SN}}
  t_{\text{NS}}^* \r)^{-1}\r]^2 
}
(it is assumed that $E = 0$). Expanding the denominators in the Taylor series,
we obtain
\eqn{
  R_A = \sum_{l, l' = 1}^\infty 2^{-(l+l')}\,  \text{Tr}\, (t_{\text{SN}}
  t_{\text{NS}}^*)^{l+l'},
\label{RA}
}
which is a convenient starting point for the semiclassical treatment. 

The semiclassical expressions for $t_{\text{SN}}$ and $t_{\text{NS}}$ are of
the form~\eq{Ssem} where the trajectories~$\gamma$ now connect the respective
leads. The calculation of the traces in equation~\eq{RA} (averaged over~$k_F$)
is completely analogous to that of the preceding section. For the classical
contribution we find 
\eqn{
  R_{A, \text{cl}} = \sum_{n=1}^{N_{\text{S}}} \sum_{m=1}^{N_{\text{N}}}
  \sum_{\gamma (n,m)} \l|\mc A_\gamma \r|^2 = T_{\text{cl}},
}
where $N_{\text{S,N}}$ is the number of channels in the leads and
$T_{\text{cl}}$ is the classical transmission coefficient for electrons. It was
taken into account that the classical traces are independent of $l + l'$ ($E =
0$) and that $\sum_{l, l' = 1}^\infty 2^{-(l+l')} = 1$. This result supports
the classical argument, according to which all trajectories that reach the
superconducting lead will contribute to~$R_{A, \text{cl}}$. The quantum
correction due to the SR pairs (there is no diagonal correction) becomes
\eqn{
  R_{A, \text{SR}} = T_{\text{SR}} \sum_{l, l' = 1}^\infty \l[ (l+l')2^{l+l'}
  \r]^{-1} = T_{\text{SR}} \ln 2,
}
where $T_{\text{SR}} < 0$ is the standard SR correction to the electron
transmission coefficient~\cite{rich02}. Thus, the weak-localisation correction
is smaller in an Andreev billiard than in the normal billiard of the same
shape.

\section{Summary and conclusions}

We presented a new method which allows us to calculate traces of semiclassical
scattering matrices in the Andreev billiards. The method was applied to
determine the density of states and conductance in the chaotic cavities. The
classical contribution to these quantities was found to be in agreement with
the existing theories. The current framework made it possible to compute the
quantum corrections to the classical results. We have shown that the
weak-localisation correction in the two-probe geometry is reduced, compared to
the normal billiard of the same shape. In the closed cavity, the quantum
corrections make the density of states negative within a small distance (of the
order of the Thouless energy) from the Fermi level. If this property is a
signature of the gap in the density of states discovered by other methods, it
will be important to understand the failure of the semiclassical theory close
to the Fermi energy. Alternatively, other, so far unknown, types of quantum
corrections could compensate the negative value.

\section*{Acknowledgements}

The author thanks Marcus Bonan\c{c}a, Petr Braun, Stefan Heusler and Klaus
Richter for the helpful discussions. The work was financially supported by the
Deutsche Forschungsgemeinschaft.

\appendix

\section{Derivations for arbitrary $l$}

\subsection{Classical contribution}
\label{genl}

We start with the generalization of integral~\eq{intl2}
\mul{
  \text{Tr}\, [S(E) S^*(-E)]^l = \l(\frac w {\pi \hbar} \r)^{2l} \int_{-\hbar
  k_F}^{\hbar k_F} dp_1 dp'_1 \cdots dp_l dp'_l \\
  \times \sum_{\substack{\gamma_1, \gamma'_1,\\ \ldots, \\ \gamma_l,\gamma'_l}}
  \mc A_{\gamma_1} \mc A_{\gamma'_1}^* \cdots \mc A_{\gamma_l} \mc
  A_{\gamma'_l}^*    \exp \l( \frac i \hbar \Phi \r)  \exp \l( \frac i \hbar 2
  l E \wbar {\mc T} \r),
\label{intl}
}
where the paths, in terms of their initial and final momenta, are $\gamma_j
(p_j \leftarrow -p'_j)$ and  $\gamma'_j (p'_j \leftarrow -p_{j+1})$ ($j = 1,
\ldots, l$; $l+1 \equiv 1$), $\Phi = \sum_{j=1}^l \l(\mc S_{\gamma_j} - \mc
S_{\gamma'_j} \r)$ and $\wbar {\mc T} = (2l)^{-1} \sum_{j=1}^l \l(\mc
T_{\gamma_j} + \mc T_{\gamma'_j} \r)$.  As in~\eq{trans2}, we transform to the
new variables
\eqn{
  \left(
    \begin{array}{c}
      a_1 \\
      \vdots\\
      a_{l-1} \\
      b
    \end{array}
  \right) 
  = C 
  \left(
    \begin{array}{c}
      p_1 \\
      \vdots\\
      p_l
    \end{array}
  \right) \quad \text{and} \quad 
  \left(
      \begin{array}{c}
      a'_1 \\
      \vdots\\
      a'_{l-1} \\
      b'
    \end{array}
  \right) 
  = C 
  \left(
    \begin{array}{c}
      p'_1 \\
      \vdots\\
      p'_l
    \end{array}
  \right)
\label{trans}
}
in such a way that $b$ and $b'$ are within the manifold
\eqn{
  \left\{
    \begin{array}{c}
      p_1 = \cdots = p_l \equiv \wbar p\\
      p'_1 = \cdots = p'_l \equiv \wbar p'
    \end{array}
  \right. \quad (|p_i|, |p'_i| \leq k_F),
\label{stsp}
}
while $a_i$ and $a'_i$ are transverse to it. This means that $C$ is an
orthogonal matrix which has the property $\sum_{j=1}^l C_{ij} =  0$  ($i=1,
\ldots, l-1$). It is convenient to choose this matrix in the form
\eqn{
  C = \left[
    \begin{array}{*9r}
      \frac 1 {\sqrt {1 \cdot 2}} (1,& -1,& 0,& 0,& 0,& \ldots,& 0,& 0) \\
      \frac 1 {\sqrt {2 \cdot 3}} (1,& 1,& -2,& 0,& 0,& \ldots,& 0,& 0) \\
      \frac 1 {\sqrt {3 \cdot 4}} (1,& 1,& 1,& -3,& 0,& \ldots,& 0,& 0) \\
                                  &&&&   \cdots &&&& \\
      \frac 1 {\sqrt {(l-1)l}}    (1,& 1,& 1,& 1,& 1,& \ldots, & 1,& -(l-1)) \\
      \frac 1 {\sqrt l}           (1,& 1,& 1,& 1,& 1,& \ldots, & 1,& 1)
    \end{array}
  \right],
\label{matC}
}
where the rows are multiplied by the factors in front of them. 

The derivatives of the phase function
\eqn{
  \frac {\pd \Phi} {\pd a_k} = \sum_{j=1}^l \l( -y_{\gamma_j}^f C^T_{jk} +
  y_{\gamma'_j}^i C^T_{j+1,k} \r),
}
and similar for $\pd \Phi/ \pd a'_k$, vanish if $\gamma_1 = \cdots = \gamma_l
\equiv \wbar \gamma$ and $\gamma'_1 = \cdots = \gamma'_l \equiv \wbar \gamma'$.
In the diagonal approximation, we have $\Phi = 0$ and $\wbar \gamma = (\wbar
\gamma')^{-1}$. In the SR quantum correction, $\wbar \gamma$ and $(\wbar
\gamma')^{-1}$ form the SR pair, and the phase is $\Phi = \pm l \Delta \mc S
(\varepsilon_{\wbar \gamma})$. 

We proceed in the diagonal approximation. The non-vanishing second
deri\-vatives of $\Phi$ at the SP point are
\eqn{
  \left. \frac {\pd^2 \Phi} {\pd a_i \pd a'_k} \right|_{a_j = a'_j = 0} =
  -\frac {\pd^2 \mc S_{\wbar \gamma} (\wbar p, -\wbar p')} {\pd \wbar p\, \pd
  (-\wbar p')} D_{ik},
}
where we introduced the $(l-1) \times (l-1)$ matrix $D_{ik} \equiv \delta_{ik}
- \sum_{j=1}^l C_{i,j+1} C^T_{jk}$. Equation~\eq{matC} yields an explicit
expression $D_{ik} = \wtilde D_{ik}/\sqrt{i (i+1) k (k+1)}$ in terms of the
matrix 
\eqn{ 
\text{\scriptsize $
  \wtilde D = \left(
    \begin{array}{*9c} 
      1 \cdot 2 + 1 & - 1 \cdot 3 & 0 & 0 & 0 & \cdots & 0 & 0 & 0 \\
      1 & 2 \cdot 3 + 1 & - 2 \cdot 4 & 0 & 0 & \cdots & 0 & 0 & 0 \\
      1 & 1 & 3 \cdot 4 + 1 & - 3 \cdot 5 & 0 & \cdots & 0 & 0 & 0 \\
                      \hdotsfor{9}                               \\
      1 & 1 & 1 & 1 & 1 & \cdots  & 1 & (l-2)(l-1) + 1 & -(l-2)l \\
      l & l & l & l & l & \cdots  & l &       l        & l^2
    \end{array}
  \right).
$}
}
Its determinant $\det \wtilde D = (l!)^2$ can be computed by adding the first
column to the second column, then adding the resulting second column to the
third column and so on. Hence, the Hessian of~$\Phi$ is given by
\eqn{
  \det \Phi'' \equiv \l|
  \begin{array}{cc}
    \l( \frac {\pd^2 \Phi} {\pd a_i \pd a_k} \r)&  
    \l( \frac {\pd^2 \Phi} {\pd a_i \pd a'_k} \r)\\[.2cm]
    \l( \frac {\pd^2 \Phi} {\pd a'_i \pd a_k} \r) & \l( \frac {\pd^2 \Phi} {\pd
    a'_i \pd a'_k} \r)
  \end{array}
  \r| = \l[\frac {\pd^2 \mc S_{\wbar \gamma} (\wbar p, -\wbar p')} {\pd \wbar
  p\, \pd (-\wbar p')} \r]^{2(l-1)} l^2.
}

To complete the SP integration, 
\eqn{
  \int \l(\prod_{j=1}^l da_j da'_j \r) \exp \l( \frac i \hbar \Phi \r)
  (\cdots) \approx \frac {(2 \pi \hbar)^{l-1}} {|\det \Phi''|^{1/2}} \exp \l(
  \frac i \hbar \Phi + i \frac \pi 4 \text{sgn}\, \Phi'' \r) (\cdots),
}
we find the difference of the number of positive and negative eigenvalues of
$\Phi''$, $\text{sgn}\, \Phi''  = 0$. This can be shown as follows: if an
eigenvector of 
\(
  \left(
    \begin{array}{cc} 
      0 & A \\
      B & 0 
    \end{array}
  \right)
\)  
with an eigenvalue~$\lambda$ is 
\(
  \left(
    \begin{array}{c}
      \alpha \\ \beta
    \end{array}
  \right)
\),
then 
\(
  \left(
    \begin{array}{c}
      \alpha \\ -\beta
    \end{array}
  \right)
\)
is an eigenvector with the eigenvalue~$-\lambda$ (here $A$ and $B$ are the
square matrices and $\alpha$ and $\beta$ are the columns of equal size). After
the last two integrations over $db\, db'$ along the manifold~\eq{stsp} we end
up with equation~\eq{TrS}.

\subsection{Diagonal quantum corrections}
\label{diagcor}

Let us repeat the SP integration over $\prod_{j=1}^l da_j da'_j$ in
appendix~\ref{genl} with the exception that now $\wbar \gamma = \wbar \gamma'$,
instead of $\wbar \gamma = (\wbar \gamma')^{-1}$, is taken at the stationary
point. This choice is compatible with the diagonality condition $\Phi = 0$ and
is possible on the manifold $\wbar p = \wbar p'$. It is convenient to perform
the remaining integrations over $db\, db'$ in the transformed coordinates
$b_\pm = l^{\mp 1/2} (b \pm b')/ \sqrt 2$ such that $\Phi = 0$ is fulfilled on
the line $b_- = 0$. The subsequent averaging over $k_F$ is expected to pick up
the contribution of the neighbourhood of this line (diagonal approximation).
Therefore, we can expand $\Phi \approx (\pd \Phi / \pd b_-)\, b_-$, where $\pd
\Phi/ \pd b_- = (y^i_{\wbar \gamma} - y^f_{\wbar \gamma})/\sqrt 2$. This leads
to the result
\mul{
  \l< \text{Tr}\, [S(E) S^*(-E)]^l \r>_{\text{diag}} \approx \\ 
  \frac 1 l \l(\frac w {\pi \hbar} \r)^2 \l< \int_{-\hbar k_F}^{\hbar k_F} db_+
  \sum_{\wbar \gamma (\wbar p \leftarrow \mp \wbar p)} |\mc A_{\wbar \gamma}|^2
  \exp \l( \frac i \hbar 2 l E \mc T_{\wbar \gamma} \r) \r. \\ 
  \l. \times \int db_- \exp \l( \frac i \hbar \frac {\pm y^i_{\wbar \gamma} -
  y^f_{\wbar \gamma}} {\sqrt 2}\, b_- \r) \r>.
\label{intdiag}
}
Note that this formula includes two classes of trajectories, $\wbar \gamma
(\wbar p \leftarrow - \wbar p)$ and \mbox{$\wbar \gamma (\wbar p \leftarrow
\wbar p)$}, and, correspondingly, two signs of $y^i_{\wbar \gamma}$ in the
second exponent. The latter class is taken into account if one starts from the
equation~\eq{intl} with $\gamma_j (p_j \leftarrow p'_j)$ and  $\gamma'_j (p'_j
\leftarrow p_{j+1})$.

One can avoid explicit calculation of equation~\eq{intdiag} by comparing it
with the result~\eq{TrSd} for $l=1$, which was derived independently. It is
then quite obvious that \eq{TrSd} is valid for arbitrary~$l$.

\section{Summation over the Sieber-Richter pairs}
\label{SR}

The purpose of this section is to fill in the steps between equations
\eq{TrSSR1} and~\eq{TrSSR}. For a trajectory~$\gamma$, the average number of
self-crossings with the crossing angle between $\varepsilon$ and $\varepsilon +
d \varepsilon$ is~\cite{rich02}
\eqn{
  P_X (\varepsilon; \mc T_\gamma)\, d \varepsilon \approx \frac {\mc T_\gamma^2
  \, v_F^2} {\pi \msf A} \sin \varepsilon \l[ \frac 1 2 - 2 \frac {\mc
  T_{\text{min}} (\varepsilon)} {\mc T_\gamma} \r] d \varepsilon ,
\label{PX}
}  
where $\mc T_{\text{min}} (\varepsilon)$ is the size of the crossing region
(logarithmically dependent on~$\varepsilon$). We took into account the
correction put forward in~\cite{heus06}, according to which the $\mc
T_{\text{min}} (\varepsilon)$ contribution in~\eq{PX} is twice as large as was
previously suggested in~\cite{rich02}. The terms of higher order in $\mc
T_{\text{min}} (\varepsilon)/ \mc T_\gamma$ are neglected. It is important to
keep the contribution proportional to~$\mc T_{\text{min}} (\varepsilon)$, since
the subsequent integration over $\varepsilon$ would eliminate the leading
term~\cite{sieb01}. 

With the help of~\eq{PX}, the sum over the self-crossings in~\eq{TrSSR1} can be
reduced to the sum over the paths as $\sum_{\{\gamma (n,m)\}}\, (\cdots) \to
\int d \varepsilon \sum_{\gamma (n,m; \varepsilon)} P_X (\varepsilon; \mc
T_\gamma)\,  (\cdots)$, where the index $\gamma (n,m; \varepsilon)$ runs over
all paths having a self-crossing of angle~$\varepsilon$. Let us, for a moment,
transform the latter sum to an integral using the sum rule~\eq{sumr}. It was
noticed by the authors of~\cite{heus06} that the sum rule has a correction
linear in~$\mc T_{\text{min}} (\varepsilon)$. It can be obtained by shifting
$\mc T \mapsto \mc T - \mc T_{\text{min}} (\varepsilon)$ in the right-hand side
of~\eq{sumr}. Thus, there are two contributions of the first order in~$\mc
T_{\text{min}} (\varepsilon)$: the one coming from the leading-order term
in~\eq{PX} and the first-order correction to the sum rule and the other
resulting from the $\mc T_{\text{min}} (\varepsilon)$ term in~\eq{PX} and the
uncorrected sum rule. Integrating over $\mc T$ explicitly, one can show that
the former contribution is two times smaller than the latter and has the
opposite sign. 

The preceding argument allows, in effect, us to consider the half of the
first-order term in~\eq{PX} and transform $\sum_{\{\gamma (n,m)\}} (\cdots) \to
\sum_{\gamma (n,m)} \int d \varepsilon P_X (\varepsilon; \mc T_\gamma)
\mspace{1mu} (\cdots)$. The $\varepsilon$~integral was computed
in~\cite{sieb01} and reads
\eqn{
  \frac {v_F^2} {\pi \msf A} \int_0^\pi d \varepsilon \cos \l[ \frac {l \Delta
  \mc S \l(\varepsilon \r)} \hbar \r] \sin (\varepsilon)\, \mc T_{\text{min}}
  (\varepsilon) \approx \frac 1 {2 N l \mc T_{\text{esc}}}.
}
Then equation~\eq{TrSSR} would follow.

\bibliography{semand}

\end{document}